\documentstyle[twoside,fleqn,espcrc2]{article}

\newcommand{\AmS}{{\protect\the\textfont2
  A\kern-.1667em\lower.5ex\hbox{M}\kern-.125emS}}

\title{The fermion determinant and the chiral gauge theory on a
lattice}

\author{Sergei V. Zenkin\address{Institute for Nuclear Research of
the Russian Academy of Sciences, 60th October Anniversary Prospect
7a, 117312 Moscow, Russia}} 

\begin{document}

\begin{abstract}

Considering as an example a simple lattice ansatz for the chiral
fermion determinant, we demonstrate that even very mild violation of
gauge invariance by the determinant at finite lattice spacing leads
to the need for another scale in the full gauge theory. This new
scale is much grater than the lattice spacing and is associated with
the gauge variables.

\end{abstract}

\maketitle

{\bf 1. The aim} of this paper is to examine consequences of a mild
violation of gauge invariance by the chiral fermion determinant for
the full gauge theory. To this end we employ a simple ansatz for the
chiral fermion determinant \cite{Z} where the violation of the gauge
invariance is in fact minimized. Namely, our ansatz ensures exact
gauge invariance of the real part of the determinant at finite
lattice spacing, in the continuum limit for smooth external fields
reproduces the consistent anomalies, and for anomaly-free theories
defines gauge invariant imaginary part of the determinant. The
violation of the gauge invariance occurs only in the imaginary part
of the determinant at finite lattice spacing and, thus, is very
mild\footnote{To our knowledge among other proposals only the overlap
formula \cite{NN} and also constructed ad hoc `hybrid' formulations
\cite{BK,HS,B} possess such properties.}. We, however, shall
demonstrate that even such a mild imperfection of the determinant
leads to strong violation of the gauge invariance of the full gauge
theory, where the gauge variables become dynamical, and that to
suppress such a violation one can allow non-smoothness of the gauge
fields only on a scale which is much grater than the lattice spacing.

We should note that the necessity for the additional scale in the
context of the chiral gauge theories was discussed earlier, see, for
example, \cite{HS,B}, and \cite{Sh} for a review and more complete
references. Our consideration yields additional arguments for that.

{\bf 2. Our ansatz} is a lattice transcription of the effective
action
\begin{equation}
\Gamma[A] \equiv \ln Z[A]
= \mbox{Tr} \ln [\partial(A) \partial^{-1}(0)],
\end{equation}
where $\partial(A)$ is a chiral Dirac operator, say, $\partial(A) =
\gamma_{\mu} (\partial_{\mu} + i g A_{\mu} P_L)$, where $P_L = (1 +
\gamma_5)/2$. The ansatz is based on the observation that both
the gauge invariance of the real part of $\Gamma[A]$, $\Re
\Gamma[A]$, and the noninvariance of the imaginary part of
$\Gamma[A]$, $\Im \Gamma[A]$, can be maintained on a lattice if in
the naive lattice formulation of (1) the domain of integration over
fermion loop momenta in all diagrams is narrowed down to ${\cal{D}} =
(- \pi/(2a), \pi/(2a))^D$, i.e.\ to the $1/2^D$-th part of the
fermion Brillouin zone ${\cal{B}} = (- \pi/a, \pi/a)^D$, where $a$ is
the lattice spacing. In this case the fermion modes which lead to the
species doubling and render any theory vector-like are no longer
dangerous, for they almost decouple from smooth gauge fields, though
are still important for restoring the gauge invariance of $\Re
\Gamma[A]$.

Our ansatz realizes such a procedure, and reads as follows:
\begin{equation}
\Gamma[A] = \mbox{Tr} \, \Theta \ln [\nabla(U) \nabla^{-1}(1)],
\end{equation}
where $\nabla(U)$ is the naive lattice transcription of the Dirac
operator
\begin{eqnarray}
\nabla&&\!\!\!\!\!\!\!\!\!\!\!\!_{m n}(U) = -
\sum_{\mu} \gamma_{\mu} \frac{1}{2 a} (U_{m \: m+\hat{\mu}}
\delta_{m+\hat{\mu} \: n} \cr
&& \quad - U_{m \: m-\hat{\mu}}
\delta_{m-\hat{\mu} \: n}), \cr
U&&\!\!\!\!\!\!\!\!\!\!\!\!_{m \: m \pm \hat{\mu}} = \exp[\pm i g a
A_{\mu}(m \pm \hat{\mu}/2) P_L],
\end{eqnarray}
and $\Theta$ is the projection operator that cuts out the proper
$1/2^D$-th part of the Brillouin zone in the fermion loop integrals:
\begin{eqnarray}
\Theta_{m n}&&\!\!\!\!\!\!\!\!\!\!\!\! =
\frac{1}{V} \sum_{p \in \cal{B}} \exp[{i p (m - n) a}] \, \Theta(p) \cr
&&\!\!\!\!\!\!\!\!\!\!\!\!= \frac{1}{N^D}
\prod_{\mu} \frac {\sin [\pi (m_{\mu} - n_{\nu})/2]}{\sin [\pi
(m_{\mu} - n_{\nu})/N]},
\end{eqnarray}
where $\Theta(p)$ is equal to unity if $p_{\mu} \in \cal{D}$ (mod $2
\pi$), or to zero otherwise. $V$ is the volume and $N^D = V/a^D$ is
the number of sites of the $D$-dimensional lattice; when $V
\rightarrow \infty$, the sum $(1/V) \sum_{p \in \cal{D}} \rightarrow
\int_{\cal{D}} d^D p /(2 \pi)^D$.

Because of the presence of $\Theta$ under the trace sign, the
effective action now cannot be written in terms of the determinant of
some operator. It, however, has the following constructive
representation:
\begin{eqnarray}
\mbox{Tr}&&\!\!\!\!\!\!\!\!\!\!\!\Theta \ln [\nabla(U)
\nabla^{-1}(1)] =
\int_{0}^{1} dt \: \mbox{Tr} \{\nabla(U - 1) \cr
&& \mbox{} \times \Theta \, [\nabla(1) +t \nabla(U - 1)]^{-1}\},
\end{eqnarray}
where $t \in [0, 1]$ is a real parameter. Note, that though $\Theta$
is non-local, it brings no much problems, since it is not gauged, and
the main price is actually the integration over $t$.

{\bf 3. Sketch of basic properties of the ansatz.} The contribution
to $\Gamma[A]$ of the $n$th order in $g A$ has the form
\begin{eqnarray}
\Gamma&&\!\!\!\!\!\!\!\!\!\!\!\!_n [A] =
\int_{\cal{B}}
\frac{d^D q_1}{(2 \pi)^D}
\cdots \frac{d^D q_{n}}{(2 \pi)^D}
\delta (q_1 + \cdots + q_{n}) \cr
&& \quad \times \mbox{tr} [ g A_{\mu_1}(q_1)
\cdots g A_{\mu_{n}}(q_{n}) \:
\Gamma_{\mu_1 \cdots \mu_{n}} (q)] \cr
\Gamma&&\!\!\!\!\!\!\!\!\!\!\!\!_{\mu_1 \cdots \mu_{n}} (q)
= \int_{\cal{D}}
\frac{d^D p}{(2 \pi)^D}
\Gamma_{\mu_1 \cdots \mu_{n}} (q; p),
\end{eqnarray}
where $p$ is the fermion loop momentum. Due to the global chiral
invariance each $\Gamma_n[A]$ is decomposed into a sum of two terms
differing from each other only in the presence or absence of
$\gamma_5$ in $\Gamma_{\mu_1 \cdots
\mu_{n}} (q; p)$. The terms without $\gamma_5$ contributes to $\Re
\Gamma$. The terms with $\gamma_5$ contributes to $\Im \Gamma$.
From the properties of the traces of $\gamma$-matrices it follows
that the real part of tr$\Gamma_{\mu_1 \cdots \mu_{n}} (q; p)$ is
periodic, and its imaginary parts is antiperiodic functions of $p$ in
the domain $\cal{D}$.

These features of the functions tr$\Gamma_{\mu_1 \cdots \mu_{n}} (q;
p)$ determine the main properties of the ansatz. In particular, from
here it follows that our ansatz yields a gauge invariant result for
$\Re \Gamma[A]$, which is
\begin{equation}
\Re \Gamma[A] = \frac{1}{2^D} \Gamma_{naive}[A] = \frac{1}{2^{D/2}}
\Re \Gamma_{staggered}[A],
\end{equation}
and that the gauge non-invariance of $\Im \Gamma[A]$ has a simple
origin very similar to what one has in the continuum theory. Indeed,
making use of Ward's identities, the expression for gauge variation
of $\Gamma_n [A]$, $\delta \Gamma_n [\omega, A]$, can be presented in
terms of the differences of two momentum integrals over the domain
$\cal{D}$ with the integrands differing from each other by a shift of
the loop momentum. Due to periodicity of the real parts of the
integrands in the domain $\cal{D}$ one can make appropriate shifts of
the integration variable that results in $\delta \Re \Gamma_n
[\omega, A] = 0$ for any $n$. For the imaginary parts $\cal{D}$ is
not the period of the corresponding integrands and such shifts result
in the appearance of a kind of surface terms, which vanish when the
regulator is removed, provided the shifts are finite and the
corresponding integrals converge or at most diverge logarithmically.
For infinitesimal gauge transformation $\omega$ the quantity $\delta
\Im \Gamma_n [\omega, A]$ has the form
\begin{eqnarray}
\delta &&\!\!\!\!\!\!\!\!\!\!\!\!\Im \Gamma_n [\omega, A] =
\int_{\cal{B}}
\frac{d^D q_1}{(2 \pi)^D}
\cdots \frac{d^D q_{n-1}}{(2 \pi)^D} \: \mbox{tr}
[g \omega (-q_1 \cr && \quad - \cdots - q_{n-1}) \, g A_{\mu_1}(q_1)
\cdots g A_{\mu_{n-1}}(q_{n-1})] \cr && \quad \times \delta \Im
\Gamma_{\mu_1
\cdots \mu_{n-1}} (q),
\end{eqnarray}
and simple estimates show that
\begin{equation}
\delta \Im \Gamma_{\mu_1 \cdots \mu_{n-1}} (q) = O(a^{n-D} (q_1 +
\cdots + q_{n-1})).
\end{equation}
Hence, for smooth external fields $A$, i.e.\ such $A$, that $\lim_{a
\rightarrow 0} A_{\mu}(q) = 0$ for any finite $qa$, all
the terms $\Im \Gamma_{n > D}[A]$ are gauge invariant in the
continuum limit, while the terms $\Im \Gamma_{n \leq D}[A]$ give rise
to the anomalies.

In \cite{Z} it was shown that the ansatz reproduces correct continuum
limit for convergent contributions to $\Gamma[A]$ of any finite order
in smooth $A$, as well as consistent chiral anomalies. Thus for
smooth external fields our ansatz is almost perfect.

{\bf 4. The problem} arises for non-smooth external fields, i.e.\
when the external momenta $q$ are no longer kept finite when the
lattice spacing tends to zero. Such a situation is realized in the
full theory, where the integration is performed over the gauge
degrees of freedom as well. As it is seen from (9), at $q = O(1/a)$
the gauge variation of the contribution of the $D+1$th order to the
effective action is no longer suppressed by a power of $a$; the
feature of the effective action that was a mild imperfection for
smooth $A$ causes the serious problem for non-smooth $A$. In our
formulation the problem is soften by the fact that due to presence of
fermion modes of the opposite chirality at high momenta \cite{Z}, the
imaginary part of $\Gamma(q)$ in (6) vanishes near the boundary of
the Brillouin zone. However, due to the properties of the
trigonometric functions we expect that $\Im \Gamma(q)$ drop
sufficiently fast at $q > f/a$, where $f$ is some fraction of $\pi$.
So the region $q \leq O(f/a)$ may be still dangerous.

The estimation (9) of the gauge variation of $\Gamma[A]$ is based on
the power counting arguments and, therefore, is quite general. Thus,
we conclude that this problem is common to all imperfect formulations
of the chiral gauge theories, including, in particular, the overlap
formula \cite{NN} (see also \cite{K} for another evidence).

{\bf 5. Need for another scale.} A universal way to suppress this
gauge non-invariance is to limit the domain of changing of the
external momenta $q$ to a region determined by a new scale $b \gg a$,
such that $\lim_{b \rightarrow 0} a/b = 0$. So, now $q \leq O(1/b)$
and in the absence of anomalies the gauge non-invariance of $\Im
\Gamma[A]$ is controlled by the ratio $r = a/b \ll 1$. The question
is how much this control is effective on a finite lattice.

We can get some idea of that applying expressions (8) and (9) to the
generic case $g A_{\mu}(n + \hat{\mu}/2) = O(1/b)$ and $g
\omega(n) = O(1)$. Returning to the finite lattice, we find
\begin{equation}
\delta \Im \Gamma_n[\omega, A] = O(N^{D}_{b} r^{n-D}).
\end{equation}
where $N^{D}_{b} = V/b^D$. Although this estimate is formal\footnote{
Eqs.\ (8), (9) are based on the expansion of the effective action in
powers of $g A$, that implies $|g A_{\mu}(n + \hat{\mu}/2)|
\ll 1/b$ and $g \omega(n) \ll 1$ and corresponds to the perturbative
regime.}, it indicates that the achieving the gauge invariance in
generic case may be more difficult than that in the perturbative
regime, and that the infinite volume limit should be correlated with
the limit $r \rightarrow 0$.

Thus, even mild imperfection of the effective action leads to the
necessity for introducing a new scale to the gauge sector. This can
be done either by imposing constraints on the gauge variable measure
(see \cite{Sh} and references therein) or by defining the gauge
variables on the sublattice with the spacing $b$ \cite{HS,B} with
their subsequent interpolation to the original lattice (then
$N^{D}_{b}$ in (10) is the number of sites of such a sublattice).
So, in this case at least part of the problems of the chiral gauge
theories moves to the gauge sector.

To conclude, we note that the only known exactly gauge invariant
formulation of the lattice chiral gauge theory \cite{Sl} (employing
non-local SLAC derivative) also involves the additional scales, which
in this case are (generalized) Pauli--Villars masses. Some arguments
for the insufficiency of a single scale lattice are given as well by
the random lattice approach \cite{Ch}. So, it appears to be plausible
that a general no-go statement about impossibility to formulate
chiral gauge theory on a lattice with only one scale is hold. \\

I am grateful to the International Science Foundation and Organizing
Committee of the ``Lattice '96" for financial support, and to
Y.~Shamir and A.~A.~Slavnov for interesting discussions. This work
was partly supported by the Russian Basic Research Fund under the
grant No. 95-02-03868a.

\end{document}